\documentclass[prl,aps,amssymb,showpacs,twocolumn]{revtex4}
\usepackage{graphicx}

\begin{document}

\title{Spin accumulation from the non-Abelian Aharonov-Bohm effect}

\author{Qin Liu$^1$, Tianxing Ma$^{1}$, Shou-Cheng Zhang$^{2,1}$}

\affiliation{$^{1}$Department of Physics, Fudan University, Shanghai
200433, China\\
 $^{2}$Department of Physics, Stanford University, Stanford, California 94305}

\date{\today}

\begin{abstract}
Recently, it has been shown that the spin-orbit coupling (SOC) of
the Dresselhaus type in $[110]$ quantum wells can be mathematically
removed by a non-Abelian gauge transformation. In the presence of an
additional uniform magnetic field, such a non-Abelian gauge flux
leads to a spin accumulation at the edges of the sample, where the
relative sign of the spin accumulation between the edges can be
tuned by the sign of the Dresselhaus SOC constant. Our prediction
can be tested by Kerr measurements within the available experimental
sensitivities.
\end{abstract}

\pacs{72.25.-b, 75.47.-m, 85.75.-d}

\maketitle

Spin transport in the presence of spin-orbit coupling has attracted
great attention in the field of spintronics. The theoretical
prediction of the spin Hall effect \cite{murakami2003,sinova2004} in
both $p$ and $n$ type doped semiconductors has lead to the
experimental observation of spin accumulation at the sample
boundaries \cite{kato2004,wunderlich2005}. Spin-orbit coupling is
the main cause of spin decoherence in solids. However, recently, it
has been realized that certain types of spin-orbit coupling,
including models with equal Rashba and Dresselhaus SOCs and the
model with Dresselhaus SOC in $[110]$ quantum wells, can be
mathematically removed by a non-Abelian gauge transformation
\cite{bernevig2006}. In these systems, the spin life time is
rendered infinite at a magic wave vector, giving rise to the
Persistent Spin Helix \cite{bernevig2006,Weber2007}.
\begin{figure}[tbp]
\begin{center}
\includegraphics[width=2.0in] {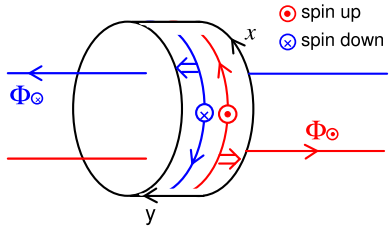}
\end{center}
\caption{(Color online) Schematic picture of Laughlin-Halperin gauge
argument in the Dresselhaus [110] model. The Dresselhaus SOC is
equivalent to two pure gauge fluxes, acting oppositely on each spin
orientation. The magnetic field points out of the cylindrical
surface where the 2DEG lives. Double arrows indicate the direction
of the Lorentz force, which leads to the spin accumulation on the
edges of the cylinder.} \label{laughlin}
\end{figure}

In this letter, we show that the equivalence of the Dresselhaus
[110] SOC with a pure non-Abelian gauge flux has another intriguing
and observable physical consequence. In the classic understanding of
the integer quantum Hall effect, Laughlin and Halperin considered a
Gedanken experiment \cite{Laughlin1981,halperin1982} in which one
adiabatically inserts a pure gauge flux through a cylinder, where
the two-dimensional electron gas (2DEG) exists on the cylindrical
surface, see Fig. \ref{laughlin}. To our knowledge, such a Gedanken
proposal has never been realized experimentally. We shall show that
the Dresselhaus [110] SOC provides an exact physical realization of
this Gedanken experiment, where the gauge flux is opposite for the
two different spin orientations. The Dresselhaus SOC $\beta$ can be
tuned continuously by changing the thickness of the [110] quantum
wells (QWs). By the Faraday's law of induction, applied separately
to both spin orientations, the adiabatic turning of the gauge flux,
or equivalently, the Dresselhaus SOC $\beta$, leads to a non-Abelian
electric field along $x$ direction \cite{jin2006},
$\mathcal{E}^x=2m\partial_t\beta(t)\sigma_z/\hbar^2$. The
non-Abelian electric field drives electrons with different
out-of-plane spin components to opposite $y$ (Hall) directions,
resulting in the spin accumulation at the sample boundaries.
Furthermore the relative sign of the spin accumulation between the
edges can be tuned by the sign of the SOC constant, and the spin
accumulation has a periodic dependence on the SOC constant, with the
equivalent Aharonov-Bohm periodicity. This effect can be detected by
Kerr measurements within the available experimental sensitivities.

Our starting point is a 2DEG system with Dresselhaus [110] SOC
interactions being turned on adiabatically at some initial time, and
a small Rashba term which we treat perturbatively. The electrons are
confined in the $x$-$y$ plane subjecting to a magnetic field
$\vec{B}=B\hat{z}, B>0$. There are two intrinsic length scales in
our systems, namely the magnetic length $l_b=\sqrt{\hbar/eB}$ and
the characteristic length of spin-orbit interaction
$l_{so}=\hbar^2/m\beta$. For a fixed magnetic field the ratio
$r=l_b/l_{so}$ plays the role of dimensionless SOC constant. We take
the periodic boundary condition in $x$ direction and the open
boundary condition in the $y$ direction. In the Landau gauge,
$\vec{A}=-By\hat{x}$, the single-particle Hamiltonian is given by
$H(r)=H_0(r)+H_R$ with
\begin{equation}
H_0=\frac{(\vec{p}+e\vec{A})^2}{2m}-\frac{2\beta}{\hbar}(p_x+eA_x)\sigma_z+\frac{g_s\mu_B
B}{2}\sigma_z+V,
\end{equation}
where $\beta$ is the Dresselhaus SOC constant, $-e$, $m$ and $\mu_B$
are respectively the charge, effective mass and Bohr magneton of the
electron, the lateral confining potential $V(y)$ vanishes for
$y\in(-L_y/2,L_y/2)$, and is infinite otherwise. In the following,
we first treat analytically the system $H_0(r)$ and then discuss the
effect of the Rashba coupling $H_R$ perturbatively.
\begin{figure}[tbp]
\begin{center}
\includegraphics[width=2.8in] {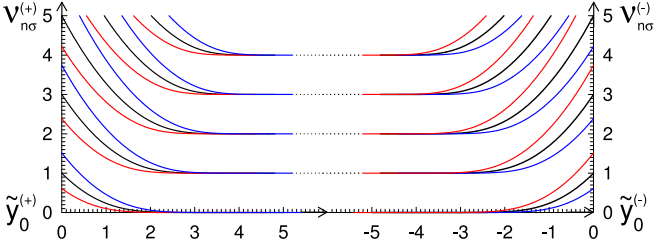}
\end{center}
\caption{(Color online) $\nu_{n\sigma}$ as a function of
$\widetilde{y}_0^{(\pm)}=(\pm L_y/2-y_0)/l_b$, the orbital center
measured from the boundaries in units of $l_b$, for positive (`$+$')
and negative (`$-$') edges. The energy spectrum at $r=0$ is shown in
black, and those at $r=0.2$ for up and down spin are shown in red
and blue.} \label{eigenenergy}
\end{figure}

For the system $H_0$ we observe that good quantum numbers are the
eigenvalues $p_x=\hbar k$ and $\sigma_z=\sigma=\pm 1$, therefore
$H_0$ is diagonalized automatically and we seek solutions of the
form $\psi(x,y,r)=\frac{1}{\sqrt{2\pi}}e^{ikx}\phi_{nk}(y,r)$ where
$\phi_{nk}$ is a two-component spinor, obeying the one dimensional
(1d) Schr$\ddot{\text{o}}$dinger equation
\begin{equation}
\left[-\frac{\hbar^2}{2m}\frac{d^2}{dy^2}+\frac{m\omega_c^2}{2}(y-y_{0\sigma})^2\right]\phi_{nk\sigma}
=\varepsilon_{nk\sigma}\phi_{nk\sigma}.
\end{equation}
In the above $y_{0\sigma}=y_0-2l_b\sigma r$ is the orbital center
for the spin component $\sigma$ and $y_0=l_b^2 k$ is that without
SOC. We notice that the $\beta$ term in Eq. (1) corresponds to a
pure gauge flux since it can be eliminated by a local non-Abelian
gauge transformation $U=\exp{(-i2m\beta x\sigma_z/\hbar^2)}$ with
the magnitude of the flux given by $\Phi_{\sigma}=4m\beta
L_x/\hbar^2\propto r$ for $\beta>0$. Therefore, when increasing
the flux $\Phi_{\sigma}$ or $r$ adiabatically, the orbital centers
of up and down spin components move to the negative and positive
Hall directions respectively, leading to accumulations near the
opposite edges. The eigenvalues of this system are given by
$E_{nk\sigma}=\varepsilon_{nk\sigma}-\hbar\omega_c(2r^2-g^{\ast}\sigma/4)$
where $\omega_c=eB/m$ and $g^{\ast}=g_sm/m_e$. If we denote the
Fermi energy of our system as $E_F=\hbar\omega_c\nu_F$ which
depends on the electron density $n_e$ and the magnetic field
$B=\frac{n_eh}{e(\nu_F+1/2)}$, the effective Fermi levels for
spin-$\sigma$ are then given by
$\nu_{F\sigma}=\nu_F+2r^2-1/2-g^{\ast}\sigma/4$. We assume that
the width $L_y$ between the two boundaries is large enough so that
the two edges are well-separated, then in the bulk region where
$|y_{0\sigma}\pm L_y/2|\gg l_b$, the edge effect can be neglected
and $\varepsilon_{nk\sigma}=\hbar\omega_c(n+1/2)$ with integers
$n$. While in the edge region where $|y_{0\sigma}\pm L_y/2|\sim
l_b$, the confining potential must be taken into account by
setting $\phi_{nk\sigma}(y=\mp L_y/2,r)=0$. In such a case if we
still keep the form
$\varepsilon_{nk\sigma}=\hbar\omega_c[\nu_{n\sigma}(k)+1/2]$,
however, $\nu_{n\sigma}$ is not necessary integers any more, the
general solution \cite{macdonald1984} to Eq. (2) may be written as
\begin{eqnarray}
\phi_{nk\sigma}(y,r)&=e^{-\frac{(y-y_{0\sigma})^2}{2l_b^2}}(C
F\left[-\frac{\nu_{n\sigma}}{2};\frac{1}{2};\frac{(y-y_{0\sigma})^2}{l_b^2}\right]\nonumber\\
&
+D\frac{y-y_{0\sigma}}{l_b}F\left[\frac{1-\nu_{n\sigma}}{2};\frac{3}{2};\frac{(y-y_{0\sigma})^2}{l_b^2}\right])
\end{eqnarray}
where $F[a;b;z]$ is the confluent hypergeometric function of the
first kind. The constants $C$ and $D$ are fixed by requiring
$\phi_{nk\sigma}(y\rightarrow\pm\infty)=0$ when considering the
negative and positive edges separately, which leads to
$\frac{D}{C}=\pm 2\tan\left(\frac{\pi\nu_{n\sigma}}{2}\right)
\frac{\Gamma(1+\nu_{n\sigma}/2)}{\Gamma(\frac{1}{2}+\nu_{n\sigma}/2)}$,
as well as the normalization of the wavefunction. The discrete
eigenvalue spectrum at each given $r$ are then determined by finding
the zeros of $\phi_{nk\sigma}(y=\mp L_y/2,r)$. The energy spectrum
at $r=0$ and 0.2 are shown in Fig. \ref{eigenenergy}.

To discuss the spin polarization, edge charge and spin current as
well as the Hall conductance, we'll focus below on their general
dependence on the magnetic field and the adiabatical change of SOC
constant, compared in particular with the well-studied case of
$r=0$. Throughout this paper we denote the corresponding quantities
in the consideration of the positive (negative) edge by a subscript
`$+(-)$'; under such a convention the out-of-plane spin polarization
near each edge gives $\langle\sigma_z(r)\rangle=\int
d\widetilde{y}\sum_{nk}^{\nu_{
F+}}|\phi_{nk+}(\widetilde{y},r)|^2-\sum_{nk}^{\nu_{
F-}}|\phi_{nk-}(\widetilde{y},r)|^2$, where
$\widetilde{y}^{(\pm)}=(y\mp L_y/2)/l_b\in[0,\mp\infty)$ is the
position variable measured from the positive (negative) edge in
units of magnetic length. We emphasize that due to the existence of
the edges, when $\nu_{F\sigma}$ exceeds the edge energy
$\nu^{\text{edge}}_{n\sigma}$ for some LL $n$,
$\nu^{\text{edge}}_{n\sigma}$ plays the role of the Fermi level for
this LL instead, and the Fermi level difference between the spin
components is no longer $g^{\ast}/2$, (which is small in general)
but increases greatly. This feature ensures the large enhancement of
the spin polarization near the edges relative to the initial
(equilibrium) case.
\begin{figure}[tbp]
\begin{center}
\includegraphics[width=3.1in] {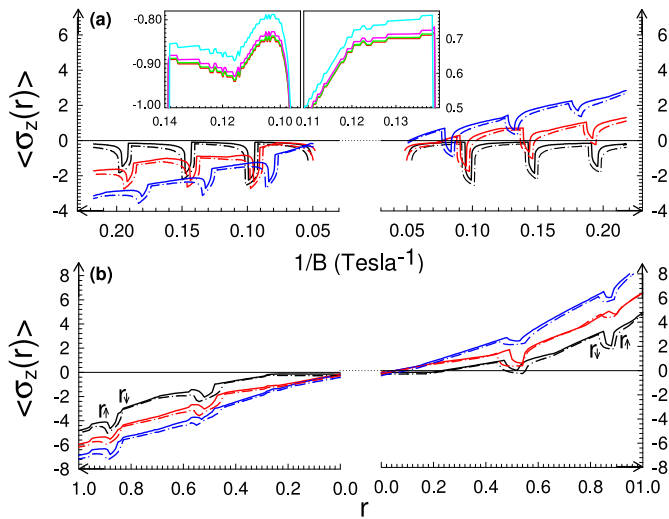}
\caption{(Color online) (a) Spin polarization as a function of
$1/B$. The results for $r=0,0.2,0.4$ are shown in black, red and
blue. Insets: Perturbed spin polarizations for $r=g^{\ast}=0.2$ at
$\delta=0.2,0.4,0.8$ are shown in green, pink and cyan. (b) Spin
polarization as a function of $r$. The results for
$\nu_F=1.0,2.0,3.0$ are shown in black, red and blue. In both
figures, right panel: negative edge; left panel: positive edge,
$n_e=5\times 10^{11}\text{cm}^{-2}$ and the solid, dash-dot lines
indicate the results for $g^{\ast}=0.2,0.4$.} \label{spinaccm}
\end{center}
\end{figure}

For $r=0$ we reproduce the well-known results for the usual 2DEG
in which the electrons are magnetized by the Zeeman interaction
with only very small negative spin polarization accumulated near
both edges. In this case the spin polarization oscillates with
$1/B$ with almost constant amplitude. This oscillation pattern is
nothing but the Shubnikov-de Hass (SdH) oscillation with the
period $\Delta_{\text{SdH}}(1/B)=1/(\Phi_0n_e)$ where $\Phi_0=h/e$
is the flux quantum. For $n_e=5\times 10^{11}\;\text{cm}^{-2}$
\cite{sih2005}, our analytic calculation gives a period of
$0.048\;(\text{T}^{-1})$ which is in precise agreement with our
numerical data. Interestingly, when the SOC constant, positive for
example, is turned on adiabatically, although the SdH oscillation
remains with the same period, its amplitude is no longer constant
but changes in different way with $1/B$ for opposite edges. Near
the negative edge the amplitude increases from negative to
positive with the decreasing of the magnetic field, while near the
positive edge the amplitude not only remains negative but also
decreases further. Moreover, as we further increase $r$, more up
(down) spins are pumped to the negative (positive) edge due to the
larger non-Abelian gauge flux. As an example, the spin
polarization for $r=0.2$ and 0.4 are shown in color in Fig.
\ref{spinaccm}(a), among which the results in red correspond to
the experimental parameters for AlGaAs [110] QWs \cite{sih2005}.
To have an overview of the dependence of the spin polarization on
the SOC constant, we also plot in Fig. \ref{spinaccm}(b) the spin
polarization as a function of $r$ at the Fermi energies
$\nu_F=1.0,2.0,3.0$ respectively, the result of which shows
clearly that the larger the SOC constant is, the more positive
(negative) the spin polarization near the negative (positive) edge
is. Furthermore, there is a similar oscillation pattern with
increasing $r$, where the turning point $r_{\sigma}$ occurs
whenever the gauge flux changes to the values so that the Fermi
energy for spin $\sigma$ crosses a bulk LL $\nu_{F\sigma}=n$, and
the period of which is given by $\Delta
r^2(\Phi_{\sigma})|_{r_{\sigma}}=1/2$. This periodicity
corresponds exactly to the Aharonov-Bohm period for both down and
up spin components of the gauge flux. For negative SOC constant
the same results are obtained with only the sign-exchange of the
spin polarization between the boundaries. The above results
present an interesting scenario for the spin accumulation:
electrons with opposite out-of-plane spin components accumulate
near the opposite edges, the relative sign of which can be tuned
by the sign of the SOC constant; by decreasing the magnetic field
or increasing the SOC constant and/or the electron density, the
magnitude of the spin polarization can be enlarged greatly
compared to the equilibrium case.

This phenomenon is a direct result of the intrinsic electric $SU(2)$
gauge field in our system. Mathematically it is clearly seen through
the expression of $y_{0\sigma}$ that for fixed $l_b$ the adiabatical
increasing of $r$ (positive) moves the orbital centers of up spin to
negative $y$ direction, and in the meanwhile carries those of down
spin to positive $y$ direction. The same conclusion is reached
through similar analysis when decreasing the magnetic field for
fixed $r$. To have an idea of the magnitude of the spin accumulation
in real materials, we recall that the range of the SOC constants
from weak to strong is \cite{bernevig2006} $10^{-13}\rightarrow
10^{-11}\;\text{eV}\cdot\text{m}$ corresponding approximately to
$r\in(0,0.5)$ which is covered by our results shown in Fig.
\ref{spinaccm}.

\begin{figure}[tbp]
\begin{center}
\includegraphics[width=1.8in] {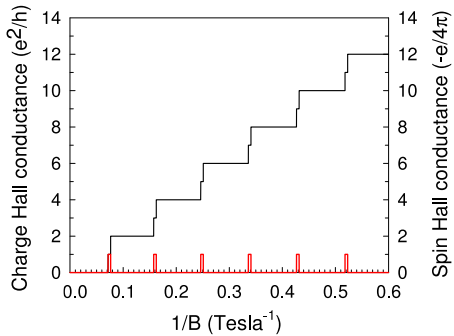}
\end{center}
\caption{(Color online) Quantized charge and spin Hall conductance
as a function of $1/B$. The CHC and SHC are shown respectively as
black and red lines. The parameters are taken as
$r=0.2,\;g^{\ast}=0.2,\;n_e=5\times 10^{11}\text{cm}^{-2}$.}
\label{hc}
\end{figure}

Let's now turn our attention to the charge and spin current as well
as their Hall conductance. The charge current operator for each spin
component is given by $j_{x\sigma}^c=-ev_{x\sigma}$ where
$v_{x\sigma}=\frac{\partial H_{0\sigma}}{\partial p_x}$ is the
velocity operator. The total charge current is the sum of that from
up and down spin components $I_x^c(r)=e\omega_cl_b\int
d\widetilde{y} \sum_{nk\sigma}^{\nu_{F\sigma}}
(\widetilde{y}+\widetilde{y}_0+2r\sigma)|\phi_{nk\sigma}|^2$, and
the charge Hall conductance (CHC) follows
$G^{c}_{xy}=G^{c}_{xy,+}+G^{c}_{xy,-}$. To compare we also consider
the spin current polarized in $z$ direction flowing along the edges
by taking the traditional definition
$j^{s}_x=\frac{\hbar}{4}\{v_x,\sigma_z\}=-(\hbar/2e) j_x^c\sigma_z$,
where we see that in our system this spin current is physically just
the $\sigma_z$-carrying charge current. Therefore the definition of
spin current is physically meaningful here in the sense that it
satisfies a well-behaved continuity equation since the charge
current does, so there is no controversy on the conservation issue
which usually can't be avoided in general SOC systems. Different
from that of charge current, the total spin current is given by the
difference of the currents from up and down spin components
$I^{s}_x(r)=I^{s}_{x+}(r)-I^{s}_{x-}(r)$ and so as the spin Hall
conductance (SHC) $G^{s}_{xy}=G^{s}_{xy,+}-G^{s}_{xy,-}$. Our
numerical results for CHC and SHC at the parameters $r=g^{\ast}=0.2$
close to those for AlGaAs [110] QWs \cite{sih2005} are shown in Fig.
\ref{hc}, it is seen that for CHC there are wider steps when it is
quantized in even multiples of the quantum of conductance, while the
steps are much narrower when it is quantized in odd multiples of the
quantum of conductance. For SHC they are non-vanishing only when the
CHC takes odd integers and the corresponding quantized number is
unity.

The above results can be easily understood by the standard
argument \cite{macdonald1984} that the charge current flows only
if there exists a chemical potential difference between the edges,
and whenever the current flows CHC is quantized in units of
$e^2/h$ with the quantized number being the number of edge states
occupied. In our system, for each spin component we have
$G^{c}_{xy,\sigma}=n+1$ with
$\nu_{F\sigma}\in(\nu^{\text{edge}}_{n-},\nu^{\text{edge}}_{n+1,-})$,
and the total CHC takes even or odd integers depending on the
relative fillings of the LLs by up and down spin. The quantization
of SHC is a direct result of the quantization of the CHC of a
spin-polarized LL in the presence of the Zeeman splitting at odd
integer filling factor. It is obvious from the relation between
the charge and spin current operators that the SHC for each spin
component is quantized in units of $-e/4\pi$ with the quantized
number equals to that of the CHC, hence it is nonvanishing only
when the total CHC takes odd integers otherwise different spin
components will cancel each other giving zero. The width of the
odd steps is $\Delta(1/B)=g^{\ast}\Delta_{\text{SdH}}/2$ which
equals approximately to $0.0048\;(\text{T}^{-1})$ for the
parameters taken in Fig. \ref{hc}. It is worth noting that in a
previous work by Bao \textit{et al.} \cite{bao2005}, similar
results for charge and (resonant) spin Hall conductance are first
obtained in a general model with both Rashba and Dresselhaus SOC,
where the SHC is not quantized and the spin accumulation decays
due to the diffusive property in these systems.

Finally we discuss the Rashba-type perturbation term
$H_R=\delta\beta[p_y\sigma_x-(\hbar k-eBy)\sigma_y]/\hbar$, where
the Rashba SOC constant has been written as $\alpha=\delta\beta$
with $\delta$ being a small perturbation number. This Hamiltonian
couples the LLs of different spin components and the wavefunction
perturbed to the first order of $H_R$ gives
\begin{equation}
\Phi_{nk\sigma}=\phi_{nk\sigma}+\sum_{n^{\prime}}\frac{W_{n^{\prime}k\overline{\sigma};nk\sigma}}{E_{nk\sigma}-E_{n^{\prime}k\overline{\sigma}}}
\phi_{n^{\prime}k\overline{\sigma}},
\end{equation}
where $E_{nk\sigma}-E_{n^{\prime}k\overline{\sigma}}=
\hbar\omega_c\left[\nu_{n\sigma}-\nu_{n^{\prime}\overline{\sigma}}+g^{\ast}(\sigma-\overline{\sigma})/4\right]$
and $W_{n^{\prime}k\overline{\sigma};nk\sigma}=\int
d\widetilde{y}\phi^{\ast}_{n^{\prime}k\overline{\sigma}}(\widetilde{y})
H_{R,\overline{\sigma}\sigma}(\widetilde{y},\partial_{\widetilde{y}})\phi_{nk\sigma}(\widetilde{y})$
with $\bar{\sigma}=-\sigma$. We have studied systematically the
effect of the perturbations to the spin polarization and some of the
results are shown as insets in Fig. \ref{spinaccm}(a). It is
concluded that the perturbations in general make the spin
polarization more positive at both edges so that
$\langle\sigma_z(r)\rangle$ is increased near negative edge while
decreased near positive edge, and the larger $\delta$ and/or $r$ is
the more the increases (decreases) are. But in general the magnitude
of these deviations is very small and will not qualitatively change
our picture of the spin accumulation.

We propose to use Kerr measurements to detect our predictions on
spin accumulation. It is calibrated in the experiment
\cite{kato2004} that for the film-like 3d sample, the spin density
of 20 Bohr magnetons per $\mu\text{m}^{-3}$ is signaled by
$1\;\mu\text{rad}$ of Kerr rotation angle. Bearing in mind that the
film thickness used in the experiment $l=0.6\;\mu\text{m}$ is 2 to 3
orders smaller than the other two dimensions, hence a similar
correspondence can be estimated for quasi-2d systems that the spin
density of 10 Bohr magnetons per $\mu\text{m}^{-2}$ lead to
$1\;\mu\text{rad}$ of Kerr rotation angle. For the parameters of
AlGaAs [110] QWs \cite{sih2005} used in our calculation, the spin
density $n_s=\langle\sigma_z\rangle/l_b^2$ is estimated to be $10^4$
Bohr magnetons per $\mu\text{m}^{-2}$ at $B=5\;\text{T}$, which is
several orders above the state-of-art detection limit in the
experiment \cite{kato2004}, and our predicted spin accumulation can
be easily detected within the available experimental sensitivities.

We wish to thank Drs Ruibao Tao, Shun-Qing Shen, Xiao-Liang Qi, Xi
Dai, Zhong Fang, Xiao-Feng Jin and Yuan Tian for helpful
discussions. This work is supported by MOE of China under the
grant number B06011, and by the NSF under grant numbers
DMR-0342832 and the US Department of Energy, Office of Basic
Energy Sciences under contract DE-AC03-76SF00515.

\bibliography{NabelABE}
\end{document}